\documentclass[12pt]{article} 
\usepackage{amsmath,amsfonts,amssymb}
\usepackage{graphicx,epic,eepic}
\usepackage[dvips]{color}
\begin{document}
%---------------------------------------------------------------------------
\title{\bf Analog model for an expanding universe}
%---------------------------------------------------------------------------
\author{Silke E. Ch. Weinfurtner\\
School of Mathematical and Computing Sciences\\
Victoria University of Wellington\\ 
PO Box 600, Wellington\\ 
New Zealand}
%------------------------------------------------
%------------------------------------------------
%---------------------------------------------------------------------------
\date{March 23, 2004; \LaTeX-ed \today}

%---------------------------------------------------------------------------
%---------------------------------------------------------------------------
%---------------------------------------------------------------------------
\maketitle
%---------------------------------------------------------------------------
%---------------------------------------------------------------------------

%---------------------------------------------------------------------------
%---------------------------------------------------------------------------
\begin{abstract}
Over the last few years numerous papers concerning analog models for gravity have been published.
It was shown that the dynamical equation of several systems (\emph{e.g.} Bose--Einstein condensates with a sink or a vortex) have the same wave equation as light in a curved-space (e.g. black holes).
In the last few months several papers were released which deal with simulations of the universe.\\

\noindent
In this article the de-Sitter universe will be compared with a freely expanding three-dimensional spherical Bose--Einstein condensate. Initially the condensate is in a harmonic trap, which suddenly will be switched off. At the same time a small perturbation will be injected in the center of the condensate cloud.

The motion of the perturbation in the expanding condensate will be discussed, and after some transformations the similarity to an expanding universe will be shown.

\begin{center}
\bigskip
\noindent
\textit{Presented at the 4th Australasian conference on General 
Relativity and Cosmology,
Monash U, Melbourne, 7--9 January 2004. }

%\bigskip
\centerline{gr-qc/0404063}

\centerline{{\sf silke.weinfurtner@mcs.vuw.ac.nz}}
\end{center}
\end{abstract}
%----------------------------------------------------------------------------

\enlargethispage{250pt}
\clearpage

\newcommand{\norm}[1]{\left\Vert#1\right\Vert}
\newcommand{\bra}[1]{\langle #1\vert}
\newcommand{\ket}[1]{\arrowvert #1 \rangle}
\newcommand{\braket}[1]{\langle #1\rangle}
\newcommand{\kin}{-\frac{\hbar^2}{2m}  \nabla^2}
\newcommand{\oppdag}[2]{\hat{#1}^\dagger(t,\vec{#2})}
\newcommand{\opp}[2]{\hat{#1}(t,\vec{#2})}
\newcommand{\bog}{\hat{\Psi}(t,\vec{x})=\Phi(t,\vec{x}) \, + \, \varepsilon \hat{\psi}(t,\vec{x})\, + \, \ldots}
\newcommand{\zeit}{i\hbar \, \frac{\partial}{\partial t}}
\newcommand{\mad}{\sqrt{\rho(t,\vec{x})}e^{i\theta(t,\vec{x})}e^{-\frac{i\mu t}{\hbar}}}

%---------------------------------------------------

%--------------------------------------------------------------
%\section{Excitations in ultra cold diluted gases in time-dependent trap}
%--------------------------------------------------------------

%----------------------------------------------------------------------
\section{Ultra-cold dilute gases}
%----------------------------------------------------------------------
The Hamiltonian for a ultra-cold gas --- consisting of $N_0$ molecules --- in a time-dependent external trap is \cite{Griffin:ab}

\begin{equation} \label{hamiltonian}
\begin{split}
\hat H=& \int{d\vec{r} \, \, \oppdag{\Psi}{r}\left[ \kin +V_{ext}(t,\vec{r})\right]\opp{\Psi}{r}} \\
       &+\frac{1}{2}\int{d\vec{r} \, d\vec{r '} \,
        \opp{\Psi}{r} \, \opp{\Psi}{r}' \, V(\vec{r}-
        \vec{r '}) \, \oppdag{\Psi}{r '} \, \oppdag{\Psi}{r}} \, . 
\end{split}
\end{equation}
The terms in the first line express the kinetic energy and the external potential,
while the second line describes the interaction part, which only takes into account two-particle interactions.
It is well known that the evolution of a field operator is given by the Heisenberg-relation:
\begin{equation}  \label{Heisenberg}
\zeit \opp{\Psi}{r}=[\opp{\Psi}{r},\hat{H}] \, .
\end{equation}
Restricting the discussion to the case of zero temperature, the ground state of the condensate can be characterized by the Hartree-Fock ansatz
\begin{equation} \label{Hartree-Fock}
	\ket{\hat{\Psi}}= \ket{\Phi} \otimes \cdots \otimes \ket{\Phi},
\end{equation}
where the field operator $\hat{\Psi}$ is replaced by $N_0$ identical macroscopic wave-functions $\Phi$,
and in that regime the interaction potential can be replaced by a so-called pseudo potential.
The model is a hard-core potential, which is completely described by the scattering length $a$. Here $a$ depends on serval factors --- for example the external field, what kind of atoms are be used and so on --- but it can experimentally measured. 
Together with the normalisation condition, the ground state of the macroscopic wave-function has to fulfill the Gross-Pitaevskii equation:
\begin{equation} \label{GPE}
\zeit \Phi(t,\vec{r})=\left( \kin + V_{ext}(t,\vec{r}) + N_0 \, g \, \vert \Phi(\vec{r}) \vert^2 - \mu \right) \Phi(t,\vec{r}) \, ,
\end{equation}
where $g=\frac{4 \pi \hbar^2}{m} \, a$ is the coupling constant.

%----------------------------------------------------------------------
\section{Sound waves in the condensate}
%----------------------------------------------------------------------
Small laser-injected density modulations in the condensate cloud had first been measured by Andrews \emph{et al.} \cite{Ketterle:1997sk}. If the perturbation is small enough the macroscopic wave-function remains in the ground state. The dynamics of the perturbation, in a time-dependent external trap, is given by the 
Gross--Pitaevskii equation.
In the eikonal approximation the density $\rho(t,\vec{r})$, and the phase $\theta(t,\vec{r})$, can be separated from each other in the following way \cite{Visser:1997ux,Barcelo:2000tg,Garay:1999sk}:
\begin{equation}
\Phi = \sqrt{\rho(t,\vec{r})} \, \exp\{i \, \theta(t,\vec{r})\}.
\end{equation}
A small perturbation in the cloud can then be characterized by a small variation in the density $\rho = \rho_0 + \varepsilon \rho_1$, and the phase $\theta = \theta_0 + \varepsilon \theta_1$.
Using the Gross--Pitaevskii equation leads to
\begin{equation} 
\begin{split}
\dot{\theta_1}=&-\frac{\hbar}{m}(\nabla \theta_0 \cdot \nabla \theta_1) - \frac{N_0 \, g}{\hbar} \rho_1 \\
\dot{\rho_1}=&-\frac{\hbar}{m}\nabla(\rho_0 \nabla \theta_1 + \rho_1 \nabla \theta_0) \, .
\end{split}
\end{equation}
Here the so called Thomas--Fermi equation has been used. In the ground state, the kinetic energy of the condensate cloud is small compared to the potential and interaction energies. 

These two equations can easily be rewritten as a single equation for the phase of the perturbation \cite{Visser:1997ux,Barcelo:2000tg,Garay:1999sk,Unruh:1980cg}:
\begin{equation}  \label{wellengleichung1}
-\ddot{\theta}_1-\partial_t(v \nabla \theta_1)-\nabla(v\dot{\theta}_1)
+\nabla\left((c^2-v^2)\nabla\theta_1\right)=0 \, .
\end{equation}
Introducing an effective metric $g_{\mu \nu}$, the differential equation for the perturbation of the phase has the same structure as the wave-equation \cite{Visser:1997ux,Barcelo:2000tg,Garay:1999sk,Unruh:1980cg}:
\begin{equation} \label{wellengleichung2}
\partial_{\mu}(\sqrt{-g} \, g^{\mu\nu} \, \partial_{\nu}\theta_{1})=0
\end{equation}
for a massless particle in a curved-space-time.
The effective metric is
\begin{equation} \label{metric}
g_{\mu \nu}=c \, \left( \begin{array}{cccc}
           -(c^2-v^2)    & -v_{x}                  & -v_{y}            & -v_{z} \\
           -v_{x}          & 1        & 0     & 0 \\
           -v_{y}          & 0           & 1   & 0  \\
           -v_{z}          & 0           & 0     & 1
 \end{array} \right).
\end{equation}
Here $c$, the speed of sound and $v$, the background velocity are given by
\begin{equation} \label{c}
c = \sqrt{\frac{g \, \rho_{0}}{m}}
\end{equation}
and
\begin{equation} \label{v}
v = \frac{\hbar}{m} \, \nabla \theta_{0} \, 
\end{equation} 
respectively.

Considering the last four equations, we see that, once the speed of sound and the background velocity are estimated, the dynamics for that configurations is fully given by wave equation (\ref{wellengleichung2}).
We now use these results to investigate a time-dependent expanding condensate.

%----------------------------------------------------------------------
\section{Expanding condensate cloud}
\begin{figure}[t]
          \begin{center}
          \input{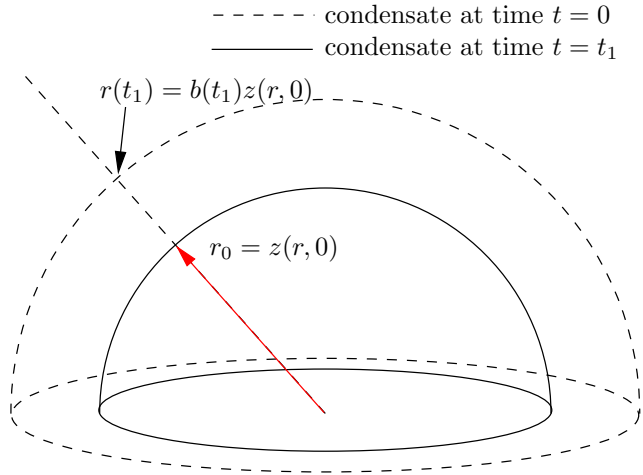}
          \caption[free expanding condensate]
          {\label{free}For times $t<0$ the condensate is in the ground state. At time $t=0$ the external trap will suddenly be switched off and the expansion starts.}
          \end{center}
          \end{figure}

%----------------------------------------------------------------------
Initially the condensate is in the ground state.
This is achieved by a static three-dimensional harmonic trap.
Suddenly, at time $t=0$ the external trap is switched off.
Immediately, the condensate cloud will expand.\\
In \cite{Castin:1996sk} it is shown, that all small portions of the gas expand along trajectories
\begin{equation} \label{r}
r(t)=z(r,t=0)\cdot b(t) \, .
\end{equation} 
Consider an infinitesimal shell at the radius $r=r_{1}$ and at the time $t=0$. After a time interval $t$ the shell radius will be increased to $r(t)=z(r_{1},t=0) \cdot b(t)$. It is convenient to choose $r(t,0)=z(r,t=0)$ corresponding the initail configuration, therefore $b(0)$ has to be unity.
In Fig.(\ref{free}) the expansion of a single shell is illustrated. \\
Then, the radial background velocity is
\begin{equation}
v_{r} = \frac{\dot{b}}{b} \, r \, .
\end{equation} 
For the speed of sound the proportionality to the density can be used, because the expansion for the volume, with respect to time, is $V(t)=b(t)^3 \cdot V(0)$. Then the speed of sound is given by
\begin{equation} \label{soundspeed}
c(t,r)=\frac{c(r)}{b(t)^{3/2}} \, .
\end{equation}  

Next, the results from this section and those from the previous section will be put together.

%----------------------------------------------------------------------
\section{de--Sitter universe}
%----------------------------------------------------------------------
In spherical coordinates the line element for the metric of Eq.(\ref{metric}) is \cite{supersilke:2003da}
\begin{equation}
ds^2 = c(r,t) \left[ \left( -c(r,t)^2 - v(r,t)^2 \right)\, dt^2
       + 2 \, v(r,t) \, dr \, dt
       + dr^2 + r^2 \, d\Omega^2 \right] \, , 
\end{equation} 
where $\Omega$ is the angular-space element.\\
Rewriting the line element in the coordinates chosen for the trajectories leads to
\begin{equation}
ds^2 = c_{0}(r) \left[ -c_{0}(t)^2 \, b(t)^{-9/2} \, dt^2 + b(t)^{1/2} \, dz^2  
       + b(t)^{1/2} \, dz^2 \, d\Omega^2 \right] \, .
\end{equation} 
Here $c_{0}(r)=c(r,t=0)$ is the speed of sound at the time $t=0$.\\

To compare this line element with the de--Sitter line element one last transformation is necessary, which eliminates the time-dependence of $dt^2$.
The appropriate transformation for that is given by
\begin{equation}
d\tau = b(t)^{-9/4} \, dt \, ,
\end{equation} 
and the resulting line element is
\begin{equation}
ds^2 = c(r) \left[ -c(r)\, d\tau^2 + b(t)^{1/2} \, r^2 \, d\Omega^2 \right] \, .
\end{equation} 
For $b(t)=(9\,H\,t + 1)^{4/9}$ --- where $H$ is the well-known Hubble constant --- the perturbation in the condensate obeys the same dynamic as the predictions for light in the de--Sitter universe.\\

\noindent
To obtain these results, two restrictions have to be made: \\
\begin{itemize}
\item The first is a restriction to time, because the Thomas--Fermi approximation --- used in order to neglect the kinetic energy --- 
becomes invalid for late times: During the expansion the potential and interaction energies will be transformed into kinetic energy. 
\item Second, the derived line element is, because of its dependence on the speed of sound, still space-dependent. But, for perturbations around the point $r=0$, the speed of sound is approximately constant. In this approximation the line element also space-independent.
\end{itemize}
Other approaches to the use of Bose--Einstein condensates to simulate the Friedmann--Robertson--Walker universe can be found in references \cite{Fedichev:2003id,Barcelo:2003et}.
%----------------------------------------------------------------------
\section{Discussion}
%----------------------------------------------------------------------
The fundamental questions are: ``Is this system realisable in the labaratory?" and "What does it teach us?" \\
In contrast to other models considered in the literature, this is one where the experimental techniques are already known. Such experiments, called \textit{time of flight measurements of expanding condensates} have been done in the laboratory. 
One main issue causing significant troubles is the instability. 
This problem could be solved using another configuration, \emph{e.g.} using a time-dependent scattering length $a(t)$, instead of an expanding condensate cloud. \\
At least, we would have a tool to simulate different scenarios for the expansion of the universe.

%------------------------------------------------------------------
%\section*{Acknowledgments}
%------------------------------------------------------------------

%------------------------------------------------------------------
%\section*{Appendix: The role of dimension}
%------------------------------------------------------------------

%----------------------------------------------------------------------

%-----------------------------------------------------------------------

%-----------------------------------------------------------------------
\end{document}